\documentclass[11pt,letterpaper]{article}

\usepackage[utf8]{inputenc}
\usepackage[letterpaper,margin=1in]{geometry}
\usepackage[skip=0.6\baselineskip plus 2pt]{parskip}

\usepackage{authblk}

\usepackage{float}
\usepackage{graphicx}
\graphicspath{{./figures/}}
\usepackage{subcaption}
\usepackage{amsmath,amssymb}
\usepackage{amsthm}
\usepackage{mathtools}
\usepackage{xcolor}
\usepackage{pgfplots}
\pgfplotsset{compat=1.17}
\usepackage{booktabs}
\usepackage{dsfont}
\usepackage{subfiles}

\usepackage[
  backend=biber,
  style=alphabetic,   
  maxnames=99,        
  maxalphanames=4,
  minalphanames=3,
  giveninits=true,
  isbn=false
]{biblatex}
\usepackage[
  colorlinks=true,
  linkcolor=blue!60!black,
  citecolor=green!45!black,
  urlcolor=blue!60!black,
  bookmarksnumbered=true
]{hyperref}
\usepackage[capitalise,nameinlink,noabbrev]{cleveref}

\usepackage{aliascnt}
\newtheorem{theorem}{Theorem}[section]      
\newcommand{\sharedtheorem}[3][plain]{
  \newaliascnt{#2}{theorem}
  \theoremstyle{#1}
  \newtheorem{#2}[#2]{#3}
  \aliascntresetthe{#2}
  \crefname{#2}{#3}{#3s}}
\sharedtheorem{lemma}{Lemma}
\sharedtheorem{proposition}{Proposition}
\sharedtheorem{corollary}{Corollary}
\sharedtheorem[definition]{definition}{Definition}
\sharedtheorem[definition]{assumption}{Assumption}
\sharedtheorem[remark]{remark}{Remark}
\theoremstyle{plain}
\crefname{theorem}{Theorem}{Theorems}

\title{At What Cost? (In)efficiency in First-Price Procurement Auctions}

\author[1]{Simone Palazzotto}
\author[1]{Andrés Cristi}

\affil[1]{
EPFL, Switzerland\par
\href{mailto:simone.palazzotto@epfl.ch}{simone.palazzotto@epfl.ch}\par
\href{mailto:andres.cristi@epfl.ch}{andres.cristi@epfl.ch}}

\date{}

\begin{document}

\pagenumbering{roman}
\maketitle
\thispagestyle{empty}

\begin{abstract}
We study the efficiency of Bayes-Nash equilibria in first-price procurement auctions. We measure efficiency by the Price of Anarchy (PoA): the
ratio between the expected social cost in equilibrium and the minimum expected
social cost resulting from a socially optimal allocation.

We first show that the standard first-price procurement auction has no
bounded PoA guarantee. We construct equilibria in which every bidder is
indifferent between all bids in a common interval, so that bids can be
arbitrarily large, and the PoA is unbounded. If costs lie in an interval
$[\underline c,\bar c]$ but bids are not restricted, the PoA equals the trivial
ratio $\bar c/\underline c$.

We hence study ways to restore a finite guarantee, first by characterizing the PoA when bids lie in $[\underline c,\bar c]$, which leads to the tight bound $1 + \frac1{e^2}(\frac{\bar c}{\underline c}-1)$. Alternatively, we introduce the notion of a \textit{phantom bidder}: the auctioneer accepts the lowest bid only if it
does not exceed a phantom bid, drawn independently of the costs, and otherwise
buys from an outside option at cost $\bar c$.

We show that a simple fixed phantom bid, i.e., a deterministic cap, gives PoA $=\frac1e\sqrt {\frac{\bar c}{\underline c}}+O(1)$. We further show that no phantom mechanism whose construction depends
only on the cost interval, under a mild regularity condition,
whatever distribution it uses and whatever it reveals about the phantom bid
before bidding, can improve on the $\sqrt{\frac{\bar c}{\underline c}}$ dependence.
\end{abstract}

\clearpage
\thispagestyle{empty}
\tableofcontents
\clearpage
\pagenumbering{arabic}

\section{Introduction}

A central question in mechanism design is how much efficiency is lost when
strategic agents play an equilibrium, compared with the outcome that a social
planner would choose. The standard measure of this loss is the \emph{Price of
Anarchy} (PoA): the ratio between the objective at the worst equilibrium and
the optimal objective \autocite{KOUTSOUPIAS200965}. Over the last two decades,
this approach has given strong worst-case guarantees for many auction formats,
including the first-price auction.

In the standard first-price auction, each bidder has a private value for the item, and 
the highest bid wins.
\textcite{10.1145/3617902} show that every Bayes--Nash equilibrium of the
first-price auction achieves at least a $1-1/e^2$ fraction of the optimal
welfare, and that this bound is tight. We extend this analysis to the reverse setting, motivated by the widespread use of this auction type.

Indeed, procurement auctions are among the most widely used auction formats in
practice. Public procurement alone accounted for $12.9\%$ of GDP on average
across OECD countries in 2021 \autocite{oecd2023glance}. Much of it, from the
construction and maintenance of public works to the purchase of goods and
services, is awarded through competitive tenders, and the most common format is
the sealed-bid first-price procurement auction: the lowest bid wins and is paid
as bid. Highway construction contracts are a prominent example, and they are
among the most studied markets in empirical work on auctions. They have been
used to study how competition affects procurement costs
\autocite{hong2002increasing}, bidding under capacity constraints
\autocite{jofre2003estimation}, bid preferences for small businesses
\autocite{krasnokutskaya2011bid}, the cost of adapting incomplete contracts
\autocite{bajari2014bidding}, and collusion
\autocite{porter1993detection,kawai2022detecting}. In these markets the
allocation matters as much as the price. Giving a contract to a supplier who is
not the most efficient wastes resources, and low winning bids can come with poor
contract performance and costly renegotiation
\autocite{decarolis2014awarding,bajari2014bidding}. The efficiency of the
allocation, measured by the social cost, is therefore a central concern.

Procurement agencies also protect themselves against expensive bids, and this
is the feature that our phantom-bidder mechanism models. An agency usually has
an outside option, such as tendering the contract again or doing the work
in-house, and it rejects bids that are too high. European procurement rules
allow contracting authorities to reject any tender whose price exceeds the
budget set before the procedure \autocite{eu2014directive}. In the construction
auctions of the Japanese national government, the maximum price is kept secret;
when no bid is below it, all bids are rejected and the suppliers bid again
\autocite{kawai2022detecting}. Whether such a maximum price should be fixed or
random, and announced or secret, are therefore practical design questions.
\cref{sec:regularized-mechanism} studies them from the point of view of
worst-case efficiency.

\subsection{Our Results}

Since the procurement setting could be roughly seen as mirroring the value auction format (costs replace values, and the lowest bid wins in place of the highest), one may hope to transfer the guarantees of the first-price auction to
procurement, but this is not immediate. Smoothness arguments, widely employed in the standard format, split the welfare
into the bidders' utilities and the seller's revenue
\autocite{10.1145/2488608.2488635,10.1145/2806883}. While, in such setting, a bid never needs to exceed the bidder's
value, in procurement, the same
split involves the auctioneer's payment and the suppliers' utilities, and these
cannot be bounded, because a supplier may submit an arbitrarily large bid. So the standard argument does not easily carry over.

In \cref{sec:PoAUnboundness}
we show that the standard first-price procurement auction, with unrestricted
bids, has no PoA guarantee. In particular, if all costs lie in an interval
$[\underline c,\bar c]$, the PoA equals the trivial bound $\bar c/\underline c$ (\cref{thm:standard-auction-PoA}). \\
\noindent Intuitively, unrestricted bids let suppliers earn arbitrarily large
utilities in equilibrium. For any fixed costs, we construct mixed equilibria in
which every supplier is indifferent between all bids in a common interval. As
the lower end of this interval grows, the differences between the suppliers'
costs become small compared with their bids. The bid distributions then become
almost identical, and every supplier wins with probability close to $1/N$,
whatever her cost. So the social cost tends to the average cost of the suppliers which yields two negative results. With unbounded costs, adding one supplier with a
very high cost makes the PoA arbitrarily large, even with a fixed number of
suppliers. With costs in $[\underline c,\bar c]$, adding many suppliers with
cost $\bar c$ makes their total winning probability close to one, while the
optimal cost stays $\underline c$, so the PoA tends to $\bar c/\underline c$.
More competition among low-cost suppliers does not help: the same ratios are
approached with any finite number of low-cost suppliers.

These pathological instances leverage the unboundedness of the bid support. The standard auction (without specifically restricting bid distributions a priori) does not control
how high the bids can go, so we
study two ways to restore a guarantee.

\begin{itemize}
    \item \textbf{Bounded bids:} in \cref{sec:firstpricetrasformation}, costs and bids
both lie in $[\underline c,\bar c]$. The change of variables
$v_i=\bar c-c_i$, $\widetilde b_i=\bar c-b_i$ maps the procurement auction
exactly to a standard first-price auction: the lowest bid becomes the highest
transformed bid, and $b_i-c_i=v_i-\widetilde b_i$. The guarantee of
\textcite{10.1145/3617902} then gives
$\mathrm{PoA}\le1+(\bar c/\underline c-1)/e^2$, and this bound is tight (\cref{thm:bounded-support}). The
bid cap is essential since bids above $\bar c$ give negative transformed bids, and
these are exactly the bids that our worst-case instance use. The guarantee is finite, but
it still grows linearly in $\bar c/\underline c$.

\item \textbf{A phantom bidder:} in \cref{sec:regularized-mechanism}, the
auctioneer keeps an outside option at cost $\bar c$ and draws a \emph{phantom
bid} $Z$, whose distribution is chosen using only the cost interval $[\underline c, \bar c]$ and not the suppliers' cost distributions. The lowest bid is accepted if it is at most
$Z$; otherwise the auctioneer uses the outside option. Equivalently, the mechanism could be seen as having an extra
fictitious bidder, the phantom bidder, taking part in the auction. Intuitively, the phantom bid
makes very high bids unprofitable in equilibrium since they most likely result in losing the auction.

There are several ways in which such cap may be enforced. The auctioneer can announce $Z$, keep it secret, or reveal partial
information about it. Let $R=\bar c/\underline c$. Our main result for this
class is that a simple optimal deterministic cap $T^\star\approx e\sqrt{\underline c\,\bar c}$, with PoA $=\sqrt R/e+(1-e^{-2})/2+o(1)$ (\cref{thm:hard-cap}) achieves the $\sqrt R$ dependence as any regular phantom mechanism that uses only the cost interval to choose the distribution of $Z$ and the signal revealed about it. In fact, we show that every mechanism in this class, under a mild regularity condition, has
$\mathrm{PoA} \ge \sqrt R/8$ (\cref{thm:signal}).
\end{itemize}
The best deterministic cap achieves worst-case PoA of order $\sqrt R$,
and no regular phantom mechanism designed using only the cost interval
can improve on this dependence. In particular, the best
deterministic cap is optimal in the class up to a constant factor.

\subsection{Technical Ideas and Related Work}
\label{sec:related}

Most PoA guarantees for auctions, including the
bound $1-1/e$ for the first-price auction, come from the \emph{smoothness}
framework\footnote{see \textcite{roughgarden2017price} for a survey}
\autocite{10.1145/2806883,roughgarden2015incomplete,10.1145/2488608.2488635}. A smoothness argument fixes,
for each bidder, a deviation that does not depend on the other bids, and shows
that the bidders' utilities from these deviations, plus a multiple of the
seller's revenue, are at least a constant fraction of the optimal welfare. In procurement, the
revenue is replaced by the auctioneer's payment, which cannot be bounded when
bids are unrestricted, and \cref{sec:PoAUnboundness} shows that no guarantee of
this kind can hold. To prove the bounds of \cref{sec:regularized-mechanism}, we instead
analyze the equilibria directly, as \textcite{10.1145/3188745.3188944} and
\textcite{10.1145/3617902} do for the first-price auction. We separate
equilibrium outcomes into allocations to the \emph{rightful winner}, the
supplier with the lowest cost, and \emph{improper allocations}, in which a
supplier with a higher cost wins. The phantom bid adds a third outcome, a
\emph{rejection}, in which the auctioneer pays $\bar c$. All our bounds come from
the trade-off between improper allocations and rejections.

Rejections are expensive, so the phantom bid must not reject cheap suppliers too
often. When bids are accepted, however, the cheapest supplier can ask for a
margin above her cost, and costlier suppliers who undercut her turn part of this
margin into improper allocations. For a deterministic cap
(\cref{sec:det-caps}), we make this trade-off exact with the reduction of
\cref{sec:firstpricetrasformation}, which maps the auction below the cap to a
first-price value auction. The worst-case instance of
\textcite{10.1145/3617902} then shows that the bound cannot be improved.

This paper falls naturally within the \textbf{PoA in auctions} literature. The smoothness framework extends from
complete information to Bayes--Nash equilibria with independent values
\autocite{roughgarden2015incomplete,syrgkanis2012bayesian}, and to several
auctions run at the same time \autocite{10.1145/2488608.2488635}. Related
analyses cover greedy and simultaneous first-price auctions
\autocite{lucier2010greedy,feldman2013simultaneous,christodoulou2016tight} and
revenue \autocite{10.1145/2600057.2602878}. For the single-item first-price
auction, \textcite{10.1145/3188745.3188944} improved the smoothness bound
$1-1/e$, and \textcite{10.1145/3617902} showed that every Bayes--Nash
equilibrium achieves a $1-1/e^2$ fraction of the optimal welfare, which is
tight, even if we consider the best equilibrium~\cite{jin2023price}.  \textcite{feldman2016correlated} study correlated and coarse correlated
equilibria of the first-price auction. 

Yet, to the best of our knowledge, the PoA of the standard
\textbf{first-price procurement auction} has not been studied. 
Related work in algorithmic mechanism design studies procurement through truthful mechanisms. The frugality literature seeks to control payments relative to benchmarks capturing the competition among suppliers
\autocite{archer2007frugal,karlin2005beyond,chen2010frugal}. 
Budget-feasible mechanism design instead seeks to maximize the buyer’s value subject to a hard constraint on total payments
\autocite{singer2010budget,bei2012budget,gravin2020optimal}. On a setting closer to ours, Babaioff et al.~\cite{babaioff2014price} studied the existence of equilibria in a price competition game between suppliers with a utility-maximizing buyer. Recently, motivated by blockchain protocols, \textcite{garimidi2026beyond} study procurement
mechanisms for a divisible good, and bound the price of
anarchy of their non-truthful versions. 

Moreover, the mechanism introduced in \cref{sec:regularized-mechanism} is basically the equivalent of putting a \textbf{reserve price} on the auction. Such caps are classically
chosen to raise revenue \autocite{myerson1981optimal,riley1981optimal}. A large
literature asks whether they should be announced or kept secret
\autocite{elyakime1994first,vincent1995bidding,rosar2014secret}, and measures
the effect of secret and random reserve prices
\autocite{li2003timber,andreyanov2022secret}. In public procurement, empirical
work studies secret reserve prices, their disclosure, and other constraints on
bids \autocite{ji2008multi,ohashi2009transparency,chassang2019collusion,kawai2022detecting}.
Bids placed by the auctioneer to enforce a reserve are known as \emph{phantom
bids} \autocite{graham1990phantom}, and \textcite{bag2000more} show that
replacing a fixed reserve by phantom bidding can raise or lower welfare. These
works mostly study revenue or the buyer's cost, with
\textcite{andreyanov2022secret} as an exception that measures surplus, and they
assume a known distribution of values or costs. We instead ask for efficiency
guarantees that hold for every distribution of costs in $[\underline c,\bar c]$
and every equilibrium. In the PoA literature, reserve prices improve the
guarantees of first-price auctions when they are close to the bidders' values,
in particular with autobidders \autocite{balseiro2021robust,colinibaldeschi2026optimal}.
Our phantom bid uses no information about the costs beyond their range.

\section{Model}
\label{sec:preliminaries}

We consider a \emph{single-item sealed-bid first-price procurement auction} with
$n$ bidders, also called suppliers. Bidder $i$ has a private cost $c_i$ for
supplying the good or service that the auctioneer wants to buy. The costs are
independent, their distributions are common knowledge, and only bidder $i$
observes $c_i$. After observing their costs, the bidders submit bids at the same
time. The bidder with the lowest bid wins and is paid her bid; the other bidders
receive nothing.

A strategy of bidder $i$ maps her cost to a bid or, for a mixed strategy, to a
distribution over bids. Our worst-case instances often give a bidder a
deterministic cost and let her randomize her bid. Fix a strategy profile. For a
bidder $i$ and a bid $b$, let $\tilde x_i(b)$ be the probability that bidder $i$
wins when she bids $b$ and the other bidders follow their strategies. If
several bidders submit the lowest bid, the auction selects one of them according
to a tie-breaking rule, which may be randomized or may favour some bidders. A bidder with cost $c_i$ who bids $b$ then has
expected utility $(b-c_i)\,\tilde x_i(b)$. The profile is a \emph{Bayes--Nash
equilibrium} if no bidder can gain by changing her bid: for every bidder $i$,
every cost $c_i$, every bid $b_i$ that she uses with this cost, and every alternative bid
$b$,
\[
(b_i-c_i)\,\tilde x_i(b_i)\ \ge\ (b-c_i)\,\tilde x_i(b).
\]
We only consider equilibria in which no bidder bids below her cost without loss since a bid below cost gives a nonpositive utility, while a bid equal
to the cost gives a nonnegative one.

Let $c^{\mathrm{eq}}$ be the cost of the winner in equilibrium, and let
$c^\star:=\min_ic_i$ be the lowest cost. The \emph{social cost} of the
equilibrium is the expected cost of the winner, and the optimal social cost is
the expected lowest cost:
\[
\mathrm{SC}:=\mathbb E\bigl[c^{\mathrm{eq}}\bigr],
\qquad
\mathrm{SC}_{\mathrm{OPT}}:=\mathbb E\bigl[c^\star\bigr].
\]
The \emph{Price of Anarchy} of the equilibrium is
$\mathrm{PoA}:=\mathrm{SC}/\mathrm{SC}_{\mathrm{OPT}}$. The PoA of an auction
format, over a class of instances, is the largest value of this ratio over all
instances in the class, all their equilibria and all tie-breaking rules. All our
upper bounds hold for every tie-breaking rule. The rule matters only for the tight instances
of \cref{rem:tightness-reverse}, in which ties at the highest price $\bar c$ (at
the cap $T$ in \cref{sec:regularized-mechanism}) are resolved in favour of the
cheapest supplier.
\section{The Standard Auction Has No Guarantee}
\label{sec:PoAUnboundness}

All results of this section come from one equilibrium construction. The idea is
that equilibrium utilities can be made arbitrarily large. As the bids grow, the
differences between the bidders' fixed costs become small compared with the
bids, and the bid distributions become almost identical. As a result, every
bidder becomes equally likely to win, whatever her cost. The following lemma
makes this precise.

\begin{lemma}
\label{lem:asymptotically-symmetric-equilibria}
Consider $N\geq 2$ bidders with deterministic costs $c_1,\ldots,c_N$, and let
$\underline c:=\min_{i\in[N]}c_i$. There is a sequence of mixed Nash equilibria
such that
$
\mathrm{SC}
\longrightarrow
\frac{1}{N}\sum_{i=1}^{N}c_i.
$
Consequently,
\[
\operatorname{PoA}
\geq
\frac{\frac{1}{N}\sum_{i=1}^{N}c_i}
     {\underline c}.
\]
\end{lemma}

\begin{proof}
Fix $L>\max_{i\in[N]}c_i$. Let bidder $i$ draw her bid from the distribution
on $[L,\infty)$ with survival function
\[
S_i(b) := 1-G_i(b)
=
\frac{b-c_i}{L-c_i}
\left(
\prod_{j=1}^{N}
\frac{L-c_j}{b-c_j}
\right)^{\frac{1}{N-1}},
\qquad b\geq L.
\]

The functions $S_i$ are chosen to make every bidder indifferent between all bids in $[L,\infty)$. If
bidder $i$ bids $b\geq L$, she wins with probability
$
\tilde x_i(b) = \prod_{j\neq i}\bigl(1-G_j(b)\bigr)
=
\frac{L-c_i}{b-c_i},
$
so her expected utility is
$
(b-c_i)\,\tilde x_i(b)
=
L-c_i,
$
which does not depend on $b$. A bid below $L$ wins with probability one, but it
gives the utility $b-c_i<L-c_i$. Hence these distributions form a mixed Nash
equilibrium. \\
Notice that these are valid distributions for a large enough value of $L$. Indeed, each $S_i$
is continuous, $S_i(L)=1$, and $S_i(b)\to0$ as $b\to\infty$, so we only need to
check that $S_i$ is decreasing. Its logarithmic derivative is
\[
\bigl(\ln S_i(b)\bigr)'
= \frac{1}{b-c_i} \left( 1-\frac{1}{N-1}\sum_{j=1}^N \frac{b-c_i}{b-c_j}\right).
\]
Each ratio $\frac{b-c_i}{b-c_j}$ tends to $1$ as $b$ grows, so the bracket tends
to $1-\frac{N}{N-1}<0$. Hence there is $L$ such that the derivative is negative
for all $b\geq L$.

Finally, write the bids as $b=Lx$, with $x\geq1$ and let $L \rightarrow \infty$ so that the lower end of the bid support is made arbitrarily large. Then
\[
1-G_i(Lx)
=
\frac{x-c_i/L}{1-c_i/L}
\left(
\prod_{j=1}^{N}
\frac{1-c_j/L}{x-c_j/L}
\right)^{\frac{1}{N-1}}
\xrightarrow{L\to\infty}
x\left(x^{-N}\right)^{\frac{1}{N-1}}
=
x^{-\frac{1}{N-1}}.
\]
So all rescaled bid distributions converge to the same continuous distribution.
The auction becomes symmetric in the limit, and each bidder wins with
probability tending to $1/N$. Hence
\[
\mathrm{SC}
=
\sum_{i=1}^{N}c_i\Pr(i\text{ wins})
\longrightarrow
\frac{1}{N}\sum_{i=1}^{N}c_i.
\]
Since the optimal social cost is $\min_{i\in[N]}c_i$, the lower bound follows.
\end{proof}

\cref{lem:asymptotically-symmetric-equilibria} gives the worst case of the
standard procurement auction. If costs are not bounded, the PoA is unbounded. If
costs are bounded, the trivial ratio between the largest and the smallest
possible cost is tight.

\begin{remark}
Our construction is related to the one-shot Bertrand folk theorem of
Baye and Morgan~\cite{baye1999folk}, who show that unbounded monopoly
profits can sustain symmetric mixed equilibria with arbitrarily large
expected profits. When all costs equal $c$, our survival functions reduce
to $(1-G_i(b))=((L-c)/(b-c))^{1/(N-1)}$, coinciding with their construction
under unit demand. Here we allow heterogeneous costs and show that,
as $L\to\infty$, each supplier's winning probability converges to $1/N$,
which yields the allocation inefficiency characterized by the following Theorem.
\end{remark}

\begin{theorem}
\label{thm:standard-auction-PoA}
The price of anarchy of the standard first-price procurement auction is
unbounded. In particular, if every bidder's cost lies in $[\underline c,\bar c]$, where
$0<\underline c<\bar c$, then
$
\operatorname{PoA}
=
\frac{\bar c}{\underline c}.
$
\end{theorem}

\begin{proof}
\emph{Bounded costs, upper bound.} The winner's cost is at most $\bar c$, and the
optimal social cost is at least $\underline c$. So every equilibrium satisfies
$
\operatorname{PoA}
\leq
\frac{\bar c}{\underline c}.
$

\emph{Bounded costs, lower bound.} Take $k$ bidders with deterministic costs
$c_1,\ldots,c_k\in[\underline c,\bar c]$, at least one of which equals
$\underline c$, and add $m$ bidders with deterministic cost $\bar c$. Applying
\cref{lem:asymptotically-symmetric-equilibria} to these $N=k+m$ bidders gives a
sequence of equilibria with
\[
\mathrm{SC}
\longrightarrow
\frac{\sum_{i=1}^{k}c_i+m\bar c}{k+m}.
\]
Every bidder becomes equally likely to win. Each high-cost bidder wins with
probability close to $1/(k+m)$, so together they win with probability close to
$m/(k+m)$. Since the optimal social cost is $\underline c$,
\[
\lim_{L\to\infty}
\frac{\mathrm{SC}}
     {\mathrm{SC}_{\mathrm{OPT}}}
=
\frac{\sum_{i=1}^{k}c_i+m\bar c}
     {(k+m)\underline c},
\qquad\text{and}\qquad
\lim_{m\to\infty}\lim_{L\to\infty}
\frac{\mathrm{SC}}
     {\mathrm{SC}_{\mathrm{OPT}}}
=
\frac{\bar c}{\underline c}.
\]

\emph{Unbounded costs.} Take one bidder with cost $\underline c>0$ and one bidder
with deterministic cost $C$. By \cref{lem:asymptotically-symmetric-equilibria},
there is a sequence of equilibria with
\[
\mathrm{SC}
\longrightarrow
\frac{\underline c+C}{2},
\qquad\text{so}\qquad
\operatorname{PoA}
\geq
\frac{\underline c+C}{2\underline c}.
\]
Since $C$ can be arbitrarily large, the price of anarchy is unbounded.
\end{proof}

\begin{remark}
More competition among low-cost bidders does not improve these guarantees. Fix
any finite number $k\ge1$ of bidders with cost $\underline c$. With bounded
costs, the trivial bound $\bar c/\underline c$ can still be approached by adding
enough high-cost bidders. With unbounded costs, a single additional bidder with
cost $C$ is enough: \cref{lem:asymptotically-symmetric-equilibria} gives
equilibria whose ratio tends to
$
\frac{k\underline c+C}{(k+1)\underline c},
$
which is unbounded as $C\to\infty$, however large $k$ is. So the inefficiency is
not caused by a lack of competition among efficient suppliers, and adding
low-cost competitors does not remove it.
\end{remark}

\section{Bounded Bids and Reduction to the Value Auction}
\label{sec:firstpricetrasformation}

The previous section shows that bounding the costs is not enough. We now also
require the bids to lie in the same interval as the costs:
$
c_i\in[\underline c,\bar c]
$
and
$
b_i\in[\underline c,\bar c].
$
This is natural when the auctioneer knows the range of the costs and does not
accept bids above $\bar c$. It does not hurt the bidders, since every bidder can
bid $b_i=c_i$ and get a nonnegative utility.

With this restriction, the procurement auction is equivalent to a standard
first-price value auction. Thus, we can apply an affine transformation from the procurement to the value auction to recover the tight characterization from \textcite{10.1145/3617902}. The following Theorem formalizes this construction.

\begin{theorem}
\label{thm:bounded-support}
Consider a first-price procurement auction with independent private costs
$c_i\in[\underline c,\bar c]$ and bids restricted to $[\underline c,\bar c]$. Let
$c^{\mathrm{eq}}$ be the cost of the winner in a Bayes--Nash equilibrium, and let
$c^\star:=\min_i c_i$. Then
$
\mathbb E[c^{\mathrm{eq}}]
\le
\left(1-\frac{1}{e^2}\right)\mathbb E[c^\star]
+
\frac{\bar c}{e^2}.
$
Consequently,
\[
\mathrm{PoA}
\le
1+\frac{1}{e^2}
\left(
\frac{\bar c}{\mathbb E[c^\star]}-1
\right)
\le
1+\frac{1}{e^2}
\left(
\frac{\bar c}{\underline c}-1
\right).
\]
\end{theorem}

\begin{proof}
Let
$
v_i:=\bar c-c_i
$
and
$
\tilde b_i:=\bar c-b_i.
$
Then $v_i,\tilde b_i\in[0,\bar c-\underline c]$. The procurement winner, who has
the lowest $b_i$, is exactly the winner of the value auction, who has the highest
$\tilde b_i$. Utilities are also preserved, since
$
(b_i-c_i)\mathbf 1\{i\text{ wins}\}
=
(v_i-\tilde b_i)\mathbf 1\{i\text{ wins}\}.
$
Therefore every Bayes--Nash equilibrium of the procurement auction gives a Bayes--Nash
equilibrium of the value auction with values $v_i$, with the same allocation. \\
Let $v^{\mathrm{eq}}$ be the value of the winner in the value auction. Since
$v^{\mathrm{eq}}=\bar c-c^{\mathrm{eq}}$ and $\max_i v_i=\bar c-c^\star$, the
guarantee of \textcite{10.1145/3617902} for the value auction gives
\[
\mathbb E[\bar c-c^{\mathrm{eq}}]
\ge
\left(1-\frac{1}{e^2}\right)
\mathbb E[\bar c-c^\star].
\]
Rearranging gives the bound on $\mathbb E[c^{\mathrm{eq}}]$. Dividing by
$\mathbb E[c^\star]$ gives the first bound on the PoA, and
$\mathbb E[c^\star]\ge\underline c$ gives the second.
\end{proof}

The bid cap $b_i\le\bar c$ is essential. Without it, the transformed bid
$\tilde b_i=\bar c-b_i$ can be negative, and the procurement auction is no longer
equivalent to a value auction with nonnegative bids. This is why, without any type of regularization, pathological
instances, like the one of \cref{sec:PoAUnboundness}, may arise.

\begin{remark}
\label{rem:tightness-reverse}
The bound in \cref{thm:bounded-support} is tight. To see this, take the
worst-case instances of the first-price value auction constructed by
\textcite[Example~4, Section~6]{10.1145/3617902}, rescaled so that values and
equilibrium bids lie in $[0,1]$. By their Lemmas~6.1 and~6.2, these are
Bayes--Nash equilibria whose optimal and equilibrium expected welfare satisfy
$
\mathbb E[v_n^\star]\longrightarrow 1
$
and
$
\mathbb E[v_n^{\mathrm{eq}}]
\longrightarrow 1-\frac{1}{e^2}.
$
Let $A:=\bar c-\underline c$, and set
$
c_i=\bar c-A v_i
$
and
$
b_i^P=\bar c-A b_i^V,
$
where $b^V$ and $b^P$ are the bids in the value and in the procurement auction.
This map preserves the allocation and multiplies utilities by $A>0$, so it
preserves equilibria. The resulting procurement instances have costs and bids in
$[\underline c,\bar c]$, and
\[
\frac{\mathbb E[c_n^{\mathrm{eq}}]}
     {\mathbb E[c_n^\star]}
=
\frac{\bar c-A\mathbb E[v_n^{\mathrm{eq}}]}
     {\bar c-A\mathbb E[v_n^\star]}
\longrightarrow
1+\frac{1}{e^2}
\left(\frac{\bar c}{\underline c}-1\right).
\]
So the bound is approached by instances with finitely many bidders.
In \cref{sec:regularized-mechanism} we use the same instances with $\bar c$
replaced by a cap $T<\bar c$.
\end{remark}

\section{Regularization Through a Phantom Bidder}
\label{sec:regularized-mechanism}

\cref{sec:PoAUnboundness,sec:firstpricetrasformation} show that the
inefficiency comes from high bids. If bids are not restricted, equilibria can
push them so high that every supplier is almost equally likely to win. If bids
are capped at $\bar c$, the PoA is finite, but it grows linearly in
$\bar c/\underline c$ (\cref{thm:bounded-support}). In this section the
auctioneer has an \emph{outside option}: she can always buy at cost $\bar c$, for
instance by doing the work in-house. She uses this option to reject expensive
bids, as procurement agencies do in practice
\autocite{eu2014directive,kawai2022detecting}. The rejection rule is given by a
\emph{phantom bid} $Z$: the lowest bid is accepted only if it is at most $Z$.

The phantom bid creates a trade-off. A supplier who asks for too much may lose, which keeps the bids low. However, every rejection costs the
auctioneer $\bar c$. The auctioneer chooses the distribution of $Z$ and what the
suppliers know about $Z$ before they bid. Let $R:=\bar c/\underline c$. Our two
main results for this class are the following.
\begin{itemize}
\item The best \emph{deterministic cap} has PoA $\sqrt R/e+O(1)$
(\cref{thm:hard-cap}).
\item No phantom mechanism, whose construction depends
only on the cost interval, whatever distribution it uses for $Z$ and whatever it
reveals about $Z$, has PoA below $\sqrt R/8$, under a mild regularity
condition (\cref{thm:signal}).
\end{itemize}
The following definition formally introduces this setup.

\begin{definition}[Phantom Bidder Construction]
\label{def:phantom-mechanism}
The auctioneer draws a pair $(Z,s)$, where $Z\le\bar c$ is the \emph{phantom bid}, and $s$ is a \emph{signal}. Its joint distribution is chosen
using only the cost interval $[\underline c,\bar c]$, and not the
suppliers' cost distributions. The pair is sampled independently of
the realized costs. The
auctioneer announces $s$, keeps $Z$ secret, and collects the bids. Let $b$ be the
lowest bid. If $b\le Z$, the bid is \emph{accepted}: its supplier is selected
and paid $b$. Otherwise, all bids are \emph{rejected}, and the auctioneer uses
the outside option, at cost $\bar c$.
\end{definition}

After seeing the signal $s$, a supplier cares about $Z$ only through the
probability that a bid $b$ is accepted,
$
S_s(b):=\Pr[Z\ge b\mid s].
$
We call $S_s$ the \emph{acceptance function}. It is nonincreasing and
left-continuous, and $S_s(b)=0$ for $b>\bar c$\footnote{This is without loss for our worst-case lower bound.
Indeed, any equilibrium under the truncated phantom bid
$\min\{Z,\bar c\}$ can be reproduced under the original phantom bid $Z$
by adding two suppliers of cost $\bar c$, both bidding $\bar c$,
with ties favouring the original suppliers. Each added supplier earns
zero and cannot profitably deviate, while bids above $\bar c$ now lose
surely. The original suppliers' payoffs are unchanged, as are the
optimal and equilibrium social costs since whenever an added supplier is
selected, her cost equals the outside-option cost $\bar c$.}. A supplier with cost $c_i$ who
bids $b$ earns $(b-c_i)\,\tilde x_i(b)\,S_s(b)$ in expectation, where
$\tilde x_i(b)$ is the probability that $b$ is the lowest bid. The social cost now
includes the outside option:
\[
\mathrm{SC}
:=\mathbb E_s\bigl[c^{\mathrm{eq}}\,\mathbf 1\{\text{accepted}\}
+\bar c\,\mathbf 1\{\text{rejected}\}\bigr],
\]
where $c^{\mathrm{eq}}$ is the cost of the supplier with the lowest bid. The
benchmark is still $\mathrm{SC}_{\mathrm{OPT}}=\mathbb E[\min_ic_i]$. The class
contains, for example, the following mechanisms:
\begin{itemize}
\item \emph{Deterministic cap} $\mathcal M_T$: $Z=T$ for a fixed number $T$.
Every bid up to $T$ is accepted, and every bid above $T$ is rejected.
\item \emph{Announced random cap} $\mathcal M_\mu$: $Z=T$ is drawn from a
distribution $\mu$ and announced ($s=T$). After the announcement, the suppliers
play $\mathcal M_T$.
\item \emph{Secret phantom} $\mathcal M^{\mathrm{sec}}_\mu$: $Z$ is drawn from
$\mu$, and nothing is announced. The lowest bid $b$ is accepted with probability
$S(b)=\mu([b,\infty))$, which decreases with $b$. Every nonincreasing,
left-continuous rule $S$ with $S(b)=0$ for $b>\bar c$ arises in this way. So this
is the same as accepting the winning bid with a probability that decreases with
the bid.
\item \emph{Partial disclosure}: for instance, $s$ says in which interval of a
fixed partition the phantom bid $Z$ lies.
\end{itemize}

For a mechanism $\mathcal M$ we consider the worst case over instances and
equilibria,
\[
\mathrm{PoA}(\mathcal M):=\sup\frac{\mathrm{SC}}{\mathrm{SC}_{\mathrm{OPT}}},
\]
where the supremum is over all instances with independent costs in
$[\underline c,\bar c]$ and all Bayes--Nash equilibria of $\mathcal M$ in which
no supplier bids below her cost. Since $\mathrm{SC}\le\bar c$ and
$\mathrm{SC}_{\mathrm{OPT}}\ge\underline c$, every phantom mechanism has
$\mathrm{PoA}(\mathcal M)\le R$.

Henceforth, we set $\underline c=1$, so that $\bar c=R$, without loss of generality, since dividing all costs, bids and the phantom bid by
$\underline c$ maps equilibria to equilibria and does not change any ratio. We also denote the first-price value auction maximum loss \cite{10.1145/3617902} of the optimal welfare by
$
\varepsilon:=\frac1{e^2}\approx0.135 .
$
In our lower bounds, the social cost exceeds the optimum for two reasons: an
\emph{improper allocation}, in which a supplier with a higher cost bids below the
cheapest supplier and wins, or a \emph{rejection}, in which the lowest bid is
above $Z$ and the auctioneer pays $R$. We first show that a simple deterministic cap achieves a sublinear dependence on $R$. Under a mild regularity condition, no other mechanism in this class improves on it by more than a constant factor.

\subsection{Deterministic caps}
\label{sec:det-caps}

For the suppliers whose cost is below $T$, a cap at $T$ brings back the setting
of \cref{sec:firstpricetrasformation}, with $\bar c$ replaced by $T$ so that no bid
above $T$ can win. The next lemma makes this precise. 

\begin{lemma}
\label{lem:capped-reduction}
Fix $T\ge1$ and a Bayes--Nash equilibrium of $\mathcal M_T$, and let
$C^\star:=\min_ic_i$. Then
\[
\mathrm{SC}\ \le\ \mathbb E\bigl[g_T(C^\star)\bigr],
\qquad\text{where}\qquad
g_T(c):=
\begin{cases}
c+\varepsilon\,(T-c), & c<T,\\
R, & c\ge T.
\end{cases}
\]
\end{lemma}

The lemma says that a cap at $T$ causes one of two losses. If the cheapest
supplier has cost $C^\star<T$, she can be accepted, and the improper allocations
waste at most the fraction $\varepsilon$ of her margin $T-C^\star$. If
$C^\star\ge T$, no bid that covers a cost is accepted, and the auctioneer pays
$R$. The proof, in \cref{app:proof-lem:capped-reduction}, maps the capped auction to a first-price
value auction with values $(T-c_i)_+$, as in the proof of
\cref{thm:bounded-support}. The only new point is that bids above $T$ are never
accepted, so a supplier with cost below $T$ gains nothing by bidding above $T$. Using this lemma we can prove the following theorem, whose full proof we defer to \cref{app:proof-thm-hard-cap}.

\begin{theorem}
\label{thm:hard-cap}
Let $T^\star:=\frac12\bigl(\sqrt{(e^2-1)^2+4e^2R}-(e^2-1)\bigr)\approx e\sqrt R$. Then
\[
\mathrm{PoA}(\mathcal{M}_{T^\star})
=\frac{\sqrt R}{e}+\frac{1-\varepsilon}{2}+O\bigl(R^{-1/2}\bigr).
\]
\end{theorem}

\begin{proof}[Proof sketch]
\emph{Upper bound.} By \cref{lem:capped-reduction},
$\mathrm{SC}/\mathrm{SC}_{\mathrm{OPT}}\le\sup_{c\in[1,R]}g_T(c)/c$. For $c<T$
the ratio $g_T(c)/c=(1-\varepsilon)+\varepsilon T/c$ decreases in $c$, so it is
at most $1+\varepsilon(T-1)$. For $c\ge T$ the ratio is $R/c\le R/T$. Hence $\mathrm{PoA}(\mathcal M_T)\le\max\{1+\varepsilon(T-1),R/T\}$; the first term increases in $T$ and the second decreases, $T^\star$ set them to be equal.

\emph{Lower bound.} Take the instances of \cref{rem:tightness-reverse}, with the
upper bound $\bar c$ on the costs replaced by $T$. They remain equilibria under
$\mathcal M_T$, and their ratios tend to $1+\varepsilon(T-1)$. Next, take one
supplier with cost slightly above $T$, together with a second supplier with cost
$R$, who bids above $R$ and is always rejected. The first supplier cannot be
accepted at any price that covers her cost, so the ratio is close to $R/T$.
\end{proof}

So, the best deterministic cap has PoA about $\sqrt R/e$. Can the auctioneer do
better by drawing the phantom bid at random, by keeping it secret, or by
revealing only part of it? In the following subsection we show that she can gain at most a
constant factor: no regular phantom mechanism beats the order $\sqrt R$.

\subsection{Every phantom mechanism}
\label{sec:signal}

Fix a phantom mechanism. The
suppliers bid after seeing the signal $s$, and from then on the phantom bid
affects them only through the acceptance function $S_s$: the lowest bid $b$
wins and is accepted with probability $S_s(b)$. So, after each signal, the
suppliers play an auction with acceptance function $S_s$, and their equilibrium
strategies depend on $S_s$. Different signals lead to different acceptance
functions, and hence to different bids. Since
the signal is public and independent of the realized costs, a strategy profile is an
equilibrium of the mechanism if and only if, for almost every signal $s$, the
bids after $s$ form an equilibrium of the auction with acceptance function
$S_s$. On a fixed instance, the social cost is the
average over $s$ of the social costs of these equilibria.

For a deterministic cap, the acceptance function is fixed, and the lower bound
of \cref{thm:hard-cap} uses instances built for it. In general, the difficulty
is to handle the acceptance function and the instance together: the cost
distributions must be fixed before $s$ is drawn, while the acceptance function,
and with it the equilibrium, changes with $s$. We handle this with the construction of \cref{lem:rival}. For fixed parameters, its cost distributions are the same for every
conditional acceptance function $S_s$, and its loss depends on the acceptance function only through the monopoly profit defined below.

\begin{definition}[Monopoly profit] Let $S$ be an acceptance function. For a cost
$c\ge1$, let
\[
\pi(c):=\max_b\,(b-c)\,S(b)
\]
be the largest expected profit of a supplier with cost $c$ whose bid is always
the lowest one. We call it her \emph{monopoly profit}. The maximum exists because
$S$ is nonincreasing and left-continuous and vanishes above $R$. Let
$b^{\mathrm m}(c)$ be the largest bid that attains it, and let
$\sigma(c):=S(b^{\mathrm m}(c))$ be the probability that this bid is accepted;
when $\pi(c)=0$ we set $\sigma(c):=0$ and $b^m(c) := R +1$. We call $T_1:=b^{\mathrm m}(1)$ the
\emph{monopoly price} of a supplier with cost $1$. For a phantom mechanism, we
write $\pi_s$, $\sigma_s$ and $b^{\mathrm m}_s$ for these quantities under
$S_s$.
\end{definition}

\begin{lemma}
\label{lem:rejections}
For every phantom mechanism $\mathcal M$ and every $c\in[1,R]$,
\[
\mathrm{PoA}(\mathcal M)\ \ge\ 1+\frac{(R-c)\,\mathbb E_s\bigl[1-\sigma_s(c)\bigr]}{c}.
\]
\end{lemma}

\begin{proof}
Take the instance with two suppliers: $M$ has cost $c$, and $D$ has cost $R$.
Since $Z\le R$, $D$ cannot earn a positive profit, so bidding above $R$ is a best
response for her, and she is then always rejected. Hence every bid of $M$ up to
$R$ is the lowest bid. After each signal $s$, $M$ bids $b^{\mathrm m}_s(c)$,
which is a best response (if $\pi_s(c)=0$, she bids above $R$ as well). She is
rejected with probability $1-\sigma_s(c)$. Hence
$\mathrm{SC}=c+(R-c)\,\mathbb E_s[1-\sigma_s(c)]$, while
$\mathrm{SC}_{\mathrm{OPT}}=c$.
\end{proof}

So a mechanism with a small PoA must accept $M$ often in these instances so that
$\sigma_s$ must be large on average. \cref{lem:envelope} in the Appendix shows that this has a price.
If suppliers with costs $t>1$ are
accepted with high probability at their monopoly prices, then a supplier with
cost $1$ earns a large monopoly profit.

To turn this profit into a loss, we need a second instance, in which the cheapest
supplier's profit leads to improper allocations. For a deterministic cap, the
instances of \cref{rem:tightness-reverse} do this. In general, however, they
would have to be adjusted to the acceptance function, and this is not possible when the instance has to be
fixed before the signal is drawn. Our explicit construction of its equilibria uses the following mild
condition.

\begin{definition}[Regularity]
\label{def:regular}
An acceptance function $S$ is \emph{regular} if $S(b)>0$ for some $b>1$, and the
profit $(b-1)\,S(b)$ of a supplier with cost $1$ is nondecreasing in $b$ on
$[1,T_1]$. A phantom mechanism is \emph{regular} if, for almost every signal
$s$, either $S_s$ is regular or $S_s(b)=0$ for all $b>1$.
\end{definition}

Regularity says that the cheapest supplier's profit increases with her price up
to her monopoly price. It holds whenever $S$ is log-concave, for instance when
$Z$ is uniform, and whenever $Z$ is regular in the sense of
\textcite{myerson1981optimal}. Every deterministic cap is regular, so every
announced random cap is a regular mechanism. If $Z$ has a log-concave density,
revealing the interval of a fixed partition that contains $Z$ also gives a
regular mechanism, because the conditional distributions are again log-concave.

The instance we use is the \emph{rival instance} $\mathcal I_n(m,\theta)$ of
\cref{lem:rival} is somehow similar to that of \cref{lem:asymptotically-symmetric-equilibria}. One supplier $M$ has cost $1$, and each of $n$ rivals has,
independently, cost $1+m$ with probability $1-\theta^{1/n}$ and cost $R$
otherwise. When $\pi$ is large compared with $m$, $M$ randomizes her price up
to $T_1$, and the present rivals sometimes undercut her, each side keeping the
other indifferent. For given $m, \theta$ and $n$, the cost distributions remain fixed across all signals and the
resulting loss depends on $S$ only through $\pi$, which the instances of
\cref{lem:rejections} control through \cref{lem:envelope}. Regularity is used
only to construct this equilibrium: $M$ can be kept indifferent over an interval
of prices only if her profit increases with her price there.\footnote{Regularity guarantees that the
rival instances have an equilibrium, which we construct explicitly. For an
arbitrary acceptance function this is not guaranteed: $S$ can jump down at many
prices, so the suppliers' payoffs are discontinuous in their bids, and standard
existence results do not apply directly. The rival instances might then have no
equilibrium for a set of signals of positive probability. They would then fall
outside the analysis of the mechanism since the PoA is a supremum over instances and
their equilibria.}

We can now show that no mechanism in this class can improve upon the sublinear
dependence. The idea is to combine the two instances above. If the PoA $L$ is small,
the instances of \cref{lem:rejections} force the mechanism to accept suppliers
with relatively low costs most of the time. By \cref{lem:envelope}, we can then bound the cheapest supplier monopoly by a parameter (function of $R$) that enters linearly the social cost bound from \cref{lem:rival}. Appropriately choosing this parameter then gives the bound.

\begin{theorem}
\label{thm:signal}
Let $\chi:=1/16$. For every $R>1$ and every regular phantom
mechanism $\mathcal M$,
\[
\mathrm{PoA}(\mathcal M)\ \ge\
L_{\mathrm{all}}(R)
:=\frac{1-\chi+\sqrt{(1-\chi)^2+\chi R}}{2}
\ \ge\ \frac{\sqrt R}{8}+\frac{15}{32}.
\]
\end{theorem}

\begin{proof}
Let $L:=\mathrm{PoA}(\mathcal M)$. We first notice that the case
$L\ge R/2$ is already covered by $L_{\mathrm{all}}$. If $R\ge2-\chi$,
a direct computation gives $L_{\mathrm{all}}(R)\le R/2\le L$.
If $R<2-\chi$, then
$L_{\mathrm{all}}(R)<L_{\mathrm{all}}(2-\chi)=1-\chi/2<1\le L$.
So let $L<R/2$, and set $X:=R/(2L)$, so that $1<X<R/L$.

For $t\in(1,X]$, \cref{lem:rejections} gives
\begin{equation}
\label{eq:thm-lemm}
\mathbb E_s\bigl[1-\sigma_s(t)\bigr]
\ \le\ \frac{(L-1)t}{R-t}
\ \le\ \frac{Lt}{R},
\end{equation}
where the second inequality holds because $t\le R/L$. Moreover,
for every signal, \cref{lem:envelope} implies
$(X-1-\pi_s)_+\le\int_1^X(1-\sigma_s(t))\,dt$.
Taking expectations and using \eqref{eq:thm-lemm} gives
\begin{equation}
\label{eq:thm-sign}
\mathbb E_s\,(X-1-\pi_s)_+
\ \le\ \int_1^X\frac{Lt}{R}\,dt
\ =\ \frac{L(X^2-1)}{2R}.
\end{equation}

Now take $\mathcal I_n(m,\frac12)$ with $m:=(X-1)/4$ from
\cref{lem:rival}. For almost every signal, \cref{lem:rival}
gives an equilibrium with
$\mathrm{SC}_s-1\ge2^{-1/n}\Phi_{m,1/2}(\pi_s)$.
The constructed strategies are measurable functions of $S_s$,
so these equilibria together form an equilibrium of $\mathcal M$.
Averaging over the signals gives
$L-1\ge2^{-1/n}\mathbb E_s\Phi_{m,1/2}(\pi_s)$.
Letting $n\to\infty$, we obtain
\begin{equation}
\label{eq:thm-up}
L-1\ \ge\ \mathbb E_s\,\Phi_{m,1/2}(\pi_s).
\end{equation}

By \cref{lem:rival}, $\Phi_{m,1/2}$ is zero up to
$2m=(X-1)/2$, and is nondecreasing and concave after that.
For $\pi>2m$, its formula is
$\Phi_{m,1/2}(\pi)=\frac m2(1-2m/\pi)$, so
$\Phi_{m,1/2}(X-1)=m/4=(X-1)/16=\chi(X-1)$.
Consequently, on $[(X-1)/2,X-1]$ it lies above the segment
joining $\bigl((X-1)/2,0\bigr)$ and
$\bigl(X-1,\chi(X-1)\bigr)$, and after $X-1$ it is at least
$\chi(X-1)$. Thus
\[
\Phi_{m,1/2}(\pi)
\ \ge\ \chi(X-1)-2\chi(X-1-\pi)_+
\qquad\text{for every }\pi\ge0.
\]

Combining \eqref{eq:thm-sign} and \eqref{eq:thm-up},
\[
L-1
\ \ge\ \chi(X-1)-2\chi\,\mathbb E_s(X-1-\pi_s)_+
\ \ge\ \chi(X-1)-\frac{\chi L(X^2-1)}{R}.
\]
At $X=R/(2L)$, the right side equals
$\chi R/(4L)-\chi+\chi L/R>\chi R/(4L)-\chi$.
Hence $L^2-(1-\chi)L-\chi R/4>0$, and $L$ is larger than
the positive root of this quadratic, which is $L_{\mathrm{all}}(R)$.
Finally, $\sqrt{(1-\chi)^2+\chi R}\ge\sqrt{\chi R}$ gives
the second inequality in the statement.
\end{proof}
\section{Conclusion}
\label{sec:conclusion}

This paper studies the efficiency of first-price procurement auctions, whose analysis cannot be immediately reduced to the selling setting. In the standard first-price procurement
auction nothing bounds the bids from above, and equilibria can push them so high
that the suppliers' costs no longer matter: the PoA is unbounded, and with costs
in $[\underline c,\bar c]$ it equals the trivial ratio $R=\bar c/\underline c$.
Capping the bids at $\bar c$ brings back the reduction to the first-price value
auction and gives the tight guarantee $1+(R-1)/e^2$, which is still linear in
$R$.

To improve upon that, we introduce the phantom-bidder construction. The auctioneer keeps an
outside option at cost $\bar c$ and accepts the lowest bid only if it does not
exceed a cap whose distribution is drawn using only the cost interval. This reduces the PoA to order $\sqrt R$, and, under a mild
regularity condition, no regular phantom mechanism can do better. 

A natural question is whether using additional information, about the number of bidders and their cost distributions, can improve this dependence. Moreover, while the phantom bidder is one way to control high bids, acting only on the lowest bid and only through acceptance, other corrective mechanisms, acting on other parts of the auction, may be implemented.

Finally, many procurement problems
involve several items, lots or bundles, and suppliers whose costs depend on the
combination of items they supply. In the selling setting, simultaneous and
combinatorial first-price auctions have constant PoA for broad classes of
valuations
\autocite{roughgarden2017price}.
These guarantees rely on the same split into utilities and revenue that fails in
procurement, and the high-bid examples of \cref{sec:PoAUnboundness} are likely
to reappear item by item. A natural direction is to use a phantom bid for each
item or bundle, and to ask whether the order $\sqrt R$ survives, how the phantom
bids of different items should be correlated, and how cost complementarities interact with
rejections. Multi-unit procurement with capacity constraints, and repeated
procurement in which the auctioneer can adapt the phantom bid over time, raise
similar questions.

\subsection*{AI Disclosure}
We used ChatGPT Plus (Astra 6.0) to formalize the construction of \cref{lem:rival}. The authors verified the correctness and originality of all content, including references.

\newpage

\printbibliography

\appendix
\crefalias{section}{appendix}
\crefalias{subsection}{subappendix}

\section*{Appendix}

\section{Proofs for Section~\ref{sec:regularized-mechanism}}
\label{app:phantom}

This appendix contains the proofs omitted from \cref{sec:regularized-mechanism}:
\cref{app:det-caps} for deterministic caps and \cref{app:signal} for the lower
bound over all phantom mechanisms. As in
\cref{sec:regularized-mechanism}, $\underline c=1$, $\bar c=R$ and
$\varepsilon=1/e^2$.

\subsection{Deterministic caps}
\label{app:det-caps}

\subsubsection{Proof of \cref{lem:capped-reduction}}\label{app:proof-lem:capped-reduction}
\begin{proof}
Fix an equilibrium of $\mathcal M_T$, and let $A:=\{C^\star<T\}$ be the event
that the cheapest supplier can be accepted.

We first show that on $A$, the lowest bid is at most $T$, and it is accepted. Let
$i$ be a supplier with $c_i<T$. If $i$ bids at most $T$, the lowest bid is at
most $T$. Suppose instead that $i$ bids above $T$. Such a bid is never accepted,
so it yields zero. A bid $b'\in(c_i,T]$ would yield at least
$(b'-c_i)\Pr[b_j>b'\ \text{for all } j\neq i]$, so in equilibrium this
probability is zero for every such $b'$. Letting $b'$ decrease to $c_i$, some
other supplier bids at most $c_i$ almost surely. In both cases the
lowest bid is at most $T$. It is accepted, because a bid equal to $Z$ is
accepted. Since no supplier bids below cost, the winner satisfies
$c_{\mathrm{win}}\le b_{\mathrm{win}}\le T$.

We can now map this capped procurement auction to an equivalent first-price value auction. Give supplier $i$ the value $v_i:=(T-c_i)_+$, and map each bid $b$ to the
transformed bid $\beta(b):=(T-b)_+$. On bids up to $T$, this is the map of the
proof of \cref{thm:bounded-support}, with $\bar c$ replaced by $T$. Every bid above $T$ is sent to the same transformed bid $0$. Such bids are
rejected in the procurement auction, while in the value auction a bid $0$ can
win a tie. We check that these ties never change an outcome or a payoff.
\begin{itemize}
\item \emph{Suppliers with $c_i\ge T$.} Their value is $0$, so they earn zero in
both auctions, whatever happens in a tie.
\item \emph{Suppliers with $c_i<T$ who bid above $T$.} By the first part of the
proof, some other supplier then bids at most $c_i<T$, that is, a transformed bid above $0$. So
they lose in both auctions.
\item \emph{Suppliers who bid exactly $T$.} The bid $T$ is also sent to $0$, so it
can tie with bids above $T$, which it beats in the procurement auction. We break
these ties in its favour. This is a valid tie-breaking rule for the value
auction: by the neighbourhood-deviation argument of \textcite{10.1145/3617902}
(Lemma~2.16), at most one supplier, say $h$, wins with the bid $T$ with positive
probability, so the rule simply gives priority to $h$ at the bid $0$.
\end{itemize}
Therefore, the two auctions have the same allocation and the same payoffs. Hence the
equilibrium of $\mathcal M_T$ gives a Bayes--Nash equilibrium of the first-price
value auction. Its welfare is $(T-c_{\mathrm{win}})\mathbf 1_A$, and its optimal
welfare is $\mathbb E[(T-C^\star)_+]$.

Finally, we know that the value auction is
$(1-\varepsilon)$-efficient \autocite{10.1145/3617902}, so
$\mathbb E[(T-c_{\mathrm{win}})\mathbf 1_A]\ge(1-\varepsilon)\,\mathbb E[(T-C^\star)_+]$.
On the complement of $A$, the social cost is at most $R$. Hence
\begin{align*}
\mathrm{SC}\ &\le\ \mathbb E\bigl[c_{\mathrm{win}}\mathbf 1_A\bigr]+R\Pr[A^c]
=\mathbb E\bigl[\bigl(T-(T-c_{\mathrm{win}})\bigr)\mathbf 1_A\bigr]+R\Pr[A^c]\\
&\le\ \mathbb E\bigl[\bigl(T-(1-\varepsilon)(T-C^\star)\bigr)\mathbf 1_A\bigr]+R\Pr[A^c],
\end{align*}
and the right side is $\mathbb E[g_T(C^\star)]$, because
$T-(1-\varepsilon)(T-c)=c+\varepsilon(T-c)$.
\end{proof}

\subsubsection{Proof of \cref{thm:hard-cap}}\label{app:proof-thm-hard-cap}
\begin{proof}
\emph{Upper bound.} Since $C^\star\ge1>0$, \cref{lem:capped-reduction}
gives
\[
\frac{\mathrm{SC}}{\mathrm{SC}_{\mathrm{OPT}}}
\le\frac{\mathbb E[g_T(C^\star)]}{\mathbb E[C^\star]}
\le\sup_{c\in[1,R]}\frac{g_T(c)}{c}.
\]
For $c<T$, $g_T(c)/c=(1-\varepsilon)+\varepsilon T/c$ decreases in $c$ and is at
most $1+\varepsilon(T-1)$. For $c\ge T$, $g_T(c)/c=R/c\le R/T$.

\emph{Lower Bound.} If $T>1$, take the instances of
\cref{rem:tightness-reverse}, with the upper bound $\bar c$ on the costs
replaced by $T$, so that all costs and bids lie in $[1,T]\subseteq[1,R]$. They
remain equilibria of $\mathcal M_T$: every bid up to $T$ is accepted, as in the
auction with bids restricted to $[1,T]$, and a deviation to a bid above $T$ is
rejected and earns zero. Their ratios therefore tend to $1+\varepsilon(T-1)$. If
$T=1$, this bound is $1$, which is trivial. If $T<R$, take one supplier with
cost $T+\delta\le R$, together with a second supplier with cost $R$. The second
supplier cannot earn a positive profit, since $Z\le R$, so she bids above $R$ and
is always rejected. The first supplier cannot be accepted at any price that
covers her cost, so the outside option is used and the ratio is
$R/(T+\delta)\to R/T$ as $\delta\downarrow0$. If $T=R$, the bound $R/T=1$ is trivial.
Thus, since the term $1+\varepsilon(T-1)$ increases and the term
$R/T$ decreases in $T$, their maximum is smallest where they are equal:
$\varepsilon T^2+(1-\varepsilon)T-R=0$, that is,
$T^2+(e^2-1)T-e^2R=0$. The positive root is $T^\star$, and the common value is
\[
1+\varepsilon(T^\star-1)=\frac{1-\varepsilon+\sqrt{(1-\varepsilon)^2+4\varepsilon R}}{2}
=\sqrt{\varepsilon R}+\frac{1-\varepsilon}{2}+O\bigl(R^{-1/2}\bigr),
\]
with $\sqrt{\varepsilon R}=\sqrt R/e$.
\end{proof}

\subsection{Every phantom mechanism}
\label{app:signal}

\begin{lemma}
\label{lem:envelope}
For every acceptance function $S$ and every $x\ge1$,
\[
\pi(1)\ \ge\ \int_1^x\sigma(t)\,dt .
\]
\end{lemma}

\begin{proof}
Let $1\le c<c'$. A supplier with cost $c$ can copy the bid $b^{\mathrm m}(c')$ of
a supplier with cost $c'$. This bid earns her
$(b^{\mathrm m}(c')-c)\,\sigma(c')=\pi(c')+(c'-c)\,\sigma(c')$, and $\pi(c)$ is
the largest profit she can earn, so
\begin{equation}
\label{eq:copy}
\pi(c)\ \ge\ \pi(c')+(c'-c)\,\sigma(c').
\end{equation}
Now take points
$1=t_0<t_1<\dots<t_k=x$ and apply \eqref{eq:copy} to each pair of consecutive
points. Adding up, and using $\pi(x)\ge0$,
\[
\pi(1)\ \ge\ \sum_{j=1}^k(t_j-t_{j-1})\,\sigma(t_j).
\]
Since $\sigma$ is nonincreasing (the optimal bidding strategy $b^m(c)$ is increasing in the cost), the right side is a lower Riemann sum of
$\int_1^x\sigma(t)\,dt$, and it converges to the integral as the points get
closer.
\end{proof}

\subsubsection{The rival instance}

\begin{lemma}
\label{lem:rival}
Let $0<m<R-1$, $0<\theta<1$ and $n\ge1$. We define the \emph{rival instance}
$\mathcal I_n(m,\theta)$, where supplier $M$ has cost $1$, so that
$\mathrm{SC}_{\mathrm{OPT}}=1$, and each of $n$ other suppliers
(the \emph{rivals}) has, independently, cost $1+m$ with probability
$1-\theta^{1/n}$ and cost $R$ otherwise. Let $S$ be an acceptance function
that is regular or satisfies $S(b)=0$ for all $b>1$, and let $\pi:=\pi(1)$.
Then the auction with acceptance function $S$ has a Bayes--Nash equilibrium
on $\mathcal I_n(m,\theta)$, without ties among real bids and without bids
below cost, such that
\[
\mathrm{SC}-\mathrm{SC}_{\mathrm{OPT}}
\ \ge\ \theta^{1/n}\,\Phi_{m,\theta}(\pi),
\qquad\text{where}\qquad
\Phi_{m,\theta}(\pi):=
\begin{cases}
m(1-\theta)\left(1-\dfrac{m}{\theta\pi}\right),
    & \theta\pi>m,\\[0.8em]
0,  & \theta\pi\le m.
\end{cases}
\]
The function $\Phi_{m,\theta}$ is nondecreasing, concave on
$[m/\theta,\infty)$, and tends to $(1-\theta)m$ as $\pi\to\infty$.
\end{lemma}

\begin{proof}
If $S(b)=0$ for all $b>1$, let $M$ bid $1$ and every rival bid $R+j$,
where $j$ is its index. No supplier can earn a positive utility by
deviating, so this is an equilibrium. Here $\pi=0$, and the loss bound
is immediate. Henceforth, assume that $S$ is regular.

We work with margins $w:=b-1$, write $g(w):=wS(1+w)$ for the profit of
$M$ at margin $w$, and let $\tau:=T_1-1$, so that $\pi=g(\tau)>0$.
By \cref{def:regular}, $g$ is nondecreasing on $[0,\tau]$. It is also
continuous there: it is left-continuous because $S$ is, and for
$0\le w<\tau$, $g(w)\le g(w^+)\le g(w)$, where the first inequality
holds because $g$ is nondecreasing and the second because $S$ is
nonincreasing.

A rival of cost $R$ bids $R+j$. Such a bid yields zero utility and is
a best response, since no bid above $R$ is accepted and no bid at or
below $R$ can give this rival a positive utility. These rivals never
win, because $M$ will always bid at most $T_1\le R$. We call a rival
\emph{present} when her cost is $1+m$.

\begin{enumerate}
\item \emph{$\tau\le m$.}
Let $M$ bid $T_1$ and every present rival bid $R+j$ as well.
Then $M$ earns her monopoly profit $\pi$. A present rival cannot win
at a price above her cost, which is at least $T_1$, so she cannot
gain by deviating. Thus the profile is an equilibrium. Since
$\theta\pi\le\theta\tau<m$, the loss bound is trivial.

\item \emph{$\tau>m$.}
We give a single construction that covers both the case where $M$
bids the rivals' cost and the case where both sides randomize.
Let $u:=\max\{g(m),\theta\pi\}$ and
$w_0:=\min\{w\in[m,\tau]:g(w)=u\}$.
The margin $w_0$ exists because $g$ is continuous and nondecreasing,
with $g(m)\le u\le\pi=g(\tau)$. For $w\in[w_0,\tau]$, let
\[
\Gamma(w):=\frac{u}{g(w)}.
\]
This function is continuous and nonincreasing, with $\Gamma(w_0)=1$
and $\Gamma(\tau)=u/\pi\ge\theta$.

We choose the rivals' strategies so that all rivals together bid a
margin above $w$ with probability   $\Gamma(w)$ on $[w_0,\tau]$.
Concretely, each present rival bids a margin above $w$ with probability
$P((b_j \ge w)\mid c_j = 1 + m) = \bigl(\Gamma(w)^{1/n}-\theta^{1/n}\bigr)/
\bigl(1-\theta^{1/n}\bigr)$ on this interval, putting any remaining
mass at the bid $R+j$. This is a valid distribution because
$\Gamma$ decreases continuously from $1$ to a value at least $\theta$.
In particular, present rivals have no atoms on $[w_0,\tau]$.

If $w_0=m$, let $M$ bid margin $m$ deterministically. Otherwise,
let $M$ bid a margin in $[w_0,\tau]$, with survival function
\[
A(w):=\tilde A(w)\,\Gamma(w)^{1/n},
\qquad
\tilde A(w):=\frac{(w_0-m)\,w}{w_0\,(w-m)}.
\]
Both factors are continuous and nonincreasing, and $A(w_0)=1$,
so this defines a valid distribution whose only atom is at $\tau$.
There are no ties among real bids since the rivals have no atoms on
$[w_0,\tau]$, and their bids above $R$ are distinct.

We now check that no supplier gains by deviating. A margin
$w\in[w_0,\tau]$ gives $M$ the utility $g(w)\Gamma(w)=u$.
A margin $w<w_0$ wins surely but yields $g(w)\le u$.
A margin $w>\tau$ wins with probability at most
$\Gamma(\tau)=u/\pi$, so it yields at most
$g(w)u/\pi\le u$. Hence $M$'s strategy is a best response.

If $w_0=m$, every present rival earns zero: a bid below $1+m$
is below her cost, and a bid above $1+m$ loses to $M$.
Thus her strategy is a best response. If $w_0>m$, a present rival
bidding margin $w\in[w_0,\tau)$ wins when $M$ and the other $n-1$
rivals bid above it. Since $\Gamma(w)S(1+w)=u/w$, her utility is
\[
u_{-M}
:=(w-m)A(w)\Gamma(w)^{(n-1)/n}S(1+w)
=(w-m)\tilde A(w)\frac{u}{w}
=\frac{u(w_0-m)}{w_0},
\]
which is independent of $w$. A margin $w\in[m,w_0)$ wins surely
and yields
$(w-m)S(1+w)=g(w)(1-m/w)\le u(1-m/w_0)$,
the same payoff. The margin $\tau$ yields at most the limit of
these payoffs from below, higher margins never win, and margins
below $m$ give a nonpositive utility. Hence the present rivals'
strategies are also best responses, and the profile is an equilibrium.

It remains to prove the loss bound. If $\theta\pi\le m$, it is
immediate. Otherwise, $g(m)\le m<\theta\pi$, so $u=\theta\pi$
and $w_0>m$. Moreover, $g(w_0)=\theta\pi\le w_0$, because
$S\le1$. The probability that $M$ bids margin $\tau$ is therefore
\[
A(\tau)
=\theta^{1/n}\frac{(w_0-m)\tau}{w_0(\tau-m)}
\ge \theta^{1/n}\left(1-\frac{m}{w_0}\right)
\ge \theta^{1/n}\left(1-\frac{m}{\theta\pi}\right).
\]

Consider the event that $M$ bids margin $\tau$ and at least one
rival is present. These events are independent, so their joint
probability is $A(\tau)(1-\theta)$. Since $\Gamma(\tau)=\theta$,
every present rival bids strictly below $\tau$. Thus, on this event,
the lowest real bid belongs to a present rival. If it is accepted,
the social cost is $1+m$; if it is rejected, the social cost is
$R>1+m$. In either case, the loss relative to the optimum is at
least $m$. Consequently,
\[
\mathrm{SC}-\mathrm{SC}_{\mathrm{OPT}}
\ge m(1-\theta)A(\tau)
\ge \theta^{1/n}m(1-\theta)
       \left(1-\frac{m}{\theta\pi}\right)
=\theta^{1/n}\Phi_{m,\theta}(\pi).
\]
\end{enumerate}

Finally, $\Phi_{m,\theta}$ is continuous at $\pi=m/\theta$.
For $\pi>m/\theta$, its derivatives are
$\Phi'_{m,\theta}(\pi)=m^2(1-\theta)/(\theta\pi^2)>0$
and $\Phi''_{m,\theta}(\pi)=-2m^2(1-\theta)/(\theta\pi^3)<0$.
These give the stated monotonicity and concavity, while the limit
$\Phi_{m,\theta}(\pi)\to(1-\theta)m$ follows directly from its formula.
\end{proof}

\end{document}